\documentclass[12pt,thmsa]{article}
\usepackage{sw20lart}
\usepackage{graphicx}
\usepackage{epsfig}
\usepackage{setspace}

\input tcilatex
\begin{document}

\title{\textbf{Extremal free energy in a simple Mean Field Theory for a Coupled
Barotropic fluid - Rotating Sphere System}}
\author{Chjan C. Lim \\
Mathematical Sciences, RPI, Troy, NY 12180, USA\\
email: limc@rpi.edu}
\maketitle

\begin{abstract}
A family of spin-lattice models are derived as convergent finite dimensional
approximations to the rest frame kinetic energy of a barotropic fluid
coupled to a massive rotating sphere. In not fixing the angular momentum of
the fluid component, there is no Hamiltonian equations of motion of the
fluid component of the coupled system. This family is used to formulate a
statistical equilibrium model for the energy - relative enstrophy theory of
the coupled barotropic fluid - rotating sphere system, known as the
spherical model, which because of its microcanonical constraint on relative
enstrophy, does not have the low temperature defect of the classical energy
- enstrophy theory. This approach differs from previous works and through
the quantum - classical mapping between quantum field theory in spatial
dimension $d$ and classical statistical mechanics in dimension $d+1,$
provides a new example of Feynman's generalization of the Least Action
Principle to problems that do not have a standard Lagrangian or Hamiltonian.
A simple mean field theory for this statistical equlibrium model is
formulated and solved, providing precise conditions on the planetary spin
and relative enstrophy in order for phase transitions to occur at positive
and negative critical temperatures, $T_{+}$ and $T_{-}.$ When the planetary
spin is relatively small, there is a single phase transition at $T_{-}<0,$
from a preferred mixed vorticity state $v=m$ for all positive temperatures
and $T<T_{-}$ to an ordered pro-rotating (west to east) flow state $v=n_u$
for $T_{-}<T<0.$ When the planetary spin is relatively large, there is an
additional phase transition at $T_{+}>0$ from a preferred mixed state $v=m$
above $T_{+}$ to an ordered counter-rotating flow state $v=n_d$ for $T<T_{+}.
$ A detailed comparison is made between the results of the mean field theory
and the results of Monte-Carlo simulations, dynamic numerical simulations
and variational theory.
\end{abstract}

\newpage\ 

\section{Introduction}

Consider the system consisting of a rotating high density rigid sphere of
radius $R,$ enveloped by a thin shell of non-divergent barotropic fluid. A
comparison with the divergent case - coupled shallow water model - will be
given in the appendix. The barotropic flow is assumed to be inviscid, apart
from an ability to exchange angular momentum and energy with the infinitely
massive solid sphere. In addition we assume that the fluid is in radiation
balance and there is no net energy gain or loss from insolation. This
provides a crude model of the complex planet - atmosphere interactions,
including the enigmatic torque mechanism responsible for the phenomenon of
atmospheric super-rotation - one of the main applications motivating this
work. We construct a simple mean field theory for the equilibrium
statistical mechanics of this problem with the specific purpose of
investigating its critical phenomenology - phase transitions that are
dependent on a few key parameters in the problem. The adjective simple
refers to the fact that in this mean field theory, a comparison of the free
energy between two special macrostates in the problem - a disordered
vorticity state and a unmixed vorticity state representing rigidly rotating
flows - is made to determine the critical temperatures. Comparisonsof the
results in this paper with those obtained by a zero temperature variational
method and also those predicted by a more sophisticated mean field theory
based on the Bragg method underscore the fact that this simple mean field
theory is nonetheless quite powerful in predicting the critical
pehenomenology of the coupled barotropic flow - rotating solid sphee system.

This paper will therefore contain new results - the simple mean field theory
and its implications - together with a fairly extensive survey of the main
ideas, approaches and results of applying statistical mechanics to complex
geophysical flows in the past two decades. The reader is invited to consult
the book \cite{limnebusbook} for the details of many topics mentioned here.

For a geophysical flow problem concerning super-rotation on a spherical
surface there is little doubt that one of the key parameters is angular
momentum of the fluid. In principle, the total angular momentum of the fluid
and solid sphere is a conserved quantity but by taking the sphere to have
infinite mass, the active part of the model is just the fluid which relaxes
by exchanging angular momentum with an infinite reservoir.

It is also clear that a quasi - 2d geophysical relaxation problem such as
this one will involve energy and enstrophy. The total or rest frame energy
of the fluid and sphere is conserved. Since we have assumed the mass of the
solid sphere to be infinite, we need only keep track of the kinetic energy
of the barotropic fluid - in the non-divergent case, there is no
gravitational potential energy in the fluid since it has uniform thickness
and density, and its upper surface is a rigid lid.

At this point there are in principle two distinct infinite reservoirs in the
relaxation problem, namely, the energy and angular momentum ones. The
author's approach to this class of geophysical problems differs from
previous works in two major aspects, namely (A) angular momentum of the
fluid is not conserved but rather the fluid relaxes by exchanging energy and
angular momentum with the solid sphere, and (B) the energy and angular
momentum reservoirs is combined into a single one. (A) is justified by one
of the aims of this approach - to model and give a relaxational explanation
of the enigmatic super-rotation problem in the Venusian atmosphere - and by
the physics of angular momentum of the coupled barotropic fluid - sphere
systems on which the approach is based. Indeed, previous works on barotropic
flows on a rotating sphere with trivial topography \cite{FS1}, \cite{Fred}
that are based on the classical energy - enstrophy theory for the barotropic
vorticity equation - hence, conserving the fluid's angular momentum - have
failed to discover any phase transitions within the valid temperature range
of the corresponding Gaussian models.

The justification of (B) takes the form of the fact - easily seen by a
calculation on the reader's part - that the rest frame kinetic energy of the
non-divergent barotropic fluid has two distinct parts, the second of which
is proportional to the fluid's changing net angular momentum relative to a
frame that is rotating at the fixed angular velocity of the sphere.

Any statistical mechanics theory on a subject matter in other than the
traditional domain of microscopic elements, requires some discussion of what
we mean by temperature and entropy. Statistical temperature for macroscopic
flows is by now a recognized scientiifc concept. Unlike the standard notion
of temperature which measures the average kinetic of molecular motions,
macroscopic flow temperature is a measure of the average kinetic energy
contained in eddies which can vary over a wide range of length scales. Given
a macroscopic flow state or macrostate vorticity distribution, there belongs
a suitable flow temperature that depends on the average energy of the eddies
in the flow.

Conservation of relative enstrophy is treated here as a microcanonical
constraint, modifying the classical energy-enstrophy theories \cite
{Kraichnan}, \cite{FS1}, \cite{Fred}, \cite{Salmon} in substantial ways,
chief amongst them being removal of the Gaussian low temperature defect
while retaining the possibility for exact solution of the model. Since the
classical energy-enstrophy theories are all doubly canonical in the energy
and enstrophy, they are essentially equivalent to Gaussian models which is
the simplest of the few exactly-solvable statistical mechanics models known
to man. Hence, previous applications of equilibrium statistical mechanics to
geophysical flows \cite{FS1}, \cite{Fred}, \cite{Salmon} have largely used
it. However, replacing the canonical constraint on enstrophy by a
microcanonical one yields a significant benefit, namely, statistical
equilibrium models, known collectively as the spherical model, that are
well-defined for all positive and negative temperatures. For the aims of
this paper, which is to investigate the precise conditions for phase
transitions in the equilibrium statistics of the coupled barotropic fluid -
rotating solid sphere system, this is important because a phase transition
could in principle, occur at any positive or negative temperature.

Higher vorticity moments are considered to be less significant than
enstrophy in statistical equilibrium models of quasi-2d geophysical flows.
We discuss this point in greater detail in \cite{Shi1}.

\section{Summary of objectives, approach and results}

It is difficult to solve the spherical model for the coupled barotropic
fluid - rotating solid sphere system analytically in closed form. Progress
on this current topic is discussed in \cite{Limsphere06}, and on the
author's webpage: www.rpi.edu/\symbol{126}limc. Other more tenable methods -
both numerical and analytical ones - will therefore have to be used to
address this problem. This research programme will use three related methods
to investigate the statistical equilibrium properties of the
energy-enstrophy theory of the coupled barotropic flows system, namely (i) a
simple mean field approach - presented here - which is based on the notion
of conserving an averaged relative enstrophy, (ii) Monte - Carlo simulations
of the spherical model (with the microcanonical constraint on relative
enstrophy), and (iii) Bragg mean field method. In this paper we present the
first. The second is presented in another paper \cite{Ding} with preliminary
results announced in the AIAA paper \cite{Lim05a}. The third is based on an
intermediate mean field method \cite{Huang} which gives a renormalized
expression for the free energy in terms of the coarse-grained barotropic
non-divergent vorticity, without using a relative enstrophy constraint \cite
{Raj}. These approximate expressions for the free energy yield values for
positive and negative critical temperatures for the coupled barotropic fluid
- rotating solid sphere system which are consistent with the mean field
results in this paper, and the Monte-carlo simulations in \cite{Ding}.
Similar to the Curie-Weiss theory for phase transitions in ferromagnets \cite
{Huang}, the special form of the hyperbolic tangent plays a key role in \cite
{Raj} and will be taken up again in a paper on the associated nonlinear
fixed point result.

The specific aim for formulating a simple mean field theory in this paper is
to find precise conditions on planetary spin, enstrophy and energy for phase
transitions in the equilibrium statistical mechanics of the coupled
barotropic fluid - rotating solid sphere system. The formulation of this
simple mean field theory is based on common features of several recent mean
field theories for the 2D Euler equations but not equivalent to any single
one of them - see Joyce and Montgomery \cite{montgomeryJoyce}, Lundgren and
Pointin \cite{Lundgren}, Miller \cite{Miller}, Robert \cite{Robert2},
Caglioti \cite{Caglioti}.

Moreover, the above mean field theories are all known to be asymptotically
exact in a nonextensive continuum limit; proofs of this property were mainly
constructed via large deviations methods and result in nonlinear elliptic
mean field equations as in \cite{Robert2}, \cite{Caglioti}. Although a
rigorous proof for this property has not yet been constructed in the case of
the coupled barotropic fluid - rotating solid sphere system, we are certain
that only the details differ from the general line of approach used in \cite
{Robert2}, \cite{Caglioti}. Our point of departure, then, is to assume that
the mean field is asymptotically exact for the coupled barotropic fluid -
rotating solid sphere system. Asymptotic exactness \cite{Caglioti}, \cite
{Miller}, \cite{montgomeryJoyce}, \cite{Lundgren} of the mean field is a
useful property for it implies that (a) a nonextensive continuum/
thermodynamic limit is well defined for the family of finite size
spin-lattice Hamiltonians $H_N$, that we will derive, as finite dimensional
approximations of the rest frame barotropic kinetic energy of nondivergent
flows, and (b) the thermal properties of this family of statistical
equilibrium spin-lattice models, in the thermodynamic limit as mesh size $N$
tends to $\infty ,$ is completely determined by the mean field.

The mean field results in this paper are obtained without solving any
nonlinear elliptic PDEs, an important point which will be discussed further,
in view of our main aim of calculating precise critical temperatures. In
particular, using this simple mean field theory, we predict at least one and
as many as two critical temperatures for the equilibrium statistical
mechanics of barotropic flows on a sphere, depending on the planetary spin $%
\Omega $ and relative enstrophy $Q_r$. For relatively low values of the
planetary spin $\Omega >0,$ there is no positive temperature phase
transition, just a negative critical temperature $T_c^{-}(\Omega ,Q_r)<0$
between a mixed vorticity state $v=m$ (without any long range order) for $T$%
, positive or negative, less hot than $T_c^{-}$ and the prograde solid-body
flow state $v=n_u$ for negative $T$ hotter than $T_c^{-}.$ For large enough
planetary spins, there is in addition a phase transition at positive
critical temperature $T_c>0$ between a hot mixed state $v=m$ and the cold
retrograde solid-body flow state $v=n_d. $

The history of applying statistical equilibrium methods to 2D turbulence is
characterized by several distinctive findings, such as Onsager's discovery
of negative temperatures when the vortices are confined, the presence of
coherent structures and an inverse cascade of energy to large scales \cite
{Kraichnan}, \cite{Fox}. Two dimensional vortex statistics also have the
dubious distinction of not supporting any phase transitions at positive
temperatures (cf. \cite{onsager}, \cite{Fox}, \cite{Kraichnan}, \cite
{montgomeryJoyce}, \cite{Miller}, \cite{Robert2}, \cite{limPhaseTransitions}%
, \cite{PS}). Until recently, quasi 2D turbulence in the context of
barotropic rotating flows over trivial topography appeared to retain many of
these properties, including the lack of phase transition. Nontrivial
topography appears to have qualitatively different and anisotropic effects.
In their classic paper \cite{Salmon}, Salmon et al derived the statistical
equilibrium states of an inviscid unforced single-layer of fluid in a
periodic box with nontrivial bottom topography in the beta plane
approximation, and reported correlations between bottom topography and
expected stream function when flow energy is low. Frederiksen \cite{Fred}
reported statistical equilibrium results for the coupled barotropic fluid -
rotating solid sphere system with nontrivial topography, that are consistent
with several numerical and experimental findings, namely, topographic
effects are asymmetric in the sense that eastward solid-body flows over
nontrivial topography is less stable compared with westward solid-body flows.

However, when the topography is trivial, Frederiksen and Sawford \cite{FS1}
reported that the eastward and westward solid-body flows are both stable but
did not find any phase transitions. By going to a simple but powerful mean
field approach, not based on solving PDEs numerically, we calculate
asymmetric positive and negative critical temperatures in terms of planetary
spin and relative enstrophy. We believe that our result on the existence of
a positive critical temperature in the rotating barotropic flows system when
the planetary spin is sufficiently large, and a negative critical
temperature for all values of the planetary spin, is the first in the large
literature on the equilibrium statistical mechanics of geophysical flows
since the papers by Salmon, Holloway and Hendershott \cite{Salmon},
Frederiksen and Sawford \cite{FS1} and Carnevale and Frederiksen \cite
{Carnevale}.

This paper is organized into the following main sections and appendix: (1)
the extended Planck's theorem, (3) derivation of Ising model-like
spin-lattice Hamiltonian for the kinetic energy of the barotropic flows, (4)
the classical energy-enstrophy theory based on this spin-lattice model, (5)
the spherical model for the barotropic flows on a rotating sphere and why it
is not exactly-solvable, (6) simple mean field theories of the spin-lattice
Hamiltonian and its thermal properties, (7) differences between the
Frederiksen and Sawford theory \cite{FS1} and the spherical model, (8)
conclusion and (9) appendix: background and properties of the coupled
barotropic fluid - rotating solid sphere system,

\section{Equilibrium statistical mechanics}

Planck's theorem will be used to find the most probable equilibrium states
in an isothermal setting: minima of a Gibbs free energy per lattice site $f$
determines the thermodynamically stable states. It is wellknown since
Onsager's seminal paper \cite{onsager} that 2D vortex statistics is
characterized by negative temperatures at high values of flow kinetic
energy. Since we are interested in exploring the entire range of flow
kinetic energy in our model, we will have to extend Planck's theorem for
negative temperatures $T$:\smallskip\ 

\textbf{Extended Planck's Theorem: }\textit{The most probable
(thermodynamically stable) state at negative (positive) temperature }$T$%
\textit{\ corresponds to maxima (minima) of the Gibbs free energy per site } 
\[
f=U-TS 
\]
\textit{where } 
\[
U=\left\langle H\right\rangle 
\]
\textit{is the internal energy per site and } 
\[
S=-k_B\int ds\text{ }p(s)\text{ }\ln p(s) 
\]
\textit{\ is the mixing entropy per site, given in terms of the probability
distribution }$p(s)$\textit{\ for site value }$s.$\textit{\ Here }$k_B$%
\textit{\ is the Boltzmann constant.}

\smallskip\ 

We sketch a proof next, using an alternative definition of the total entropy
due to Boltzmann. The total free energy $F=U-TS$ is related to the Gibbs
partition function by 
\begin{equation}
Z(\beta )=e^{-\beta F(m)}=e^{-\beta (U(m)-TS(m))}  \label{thumb}
\end{equation}
where 
\[
F(m)=U(m)-TS(m) 
\]
denotes extremal values of the free energy, 
\[
\beta =\frac 1{k_BT} 
\]
is the inverse temperature and 
\[
S(m)=k_B\ln W(m) 
\]
is the total entropy in terms of the degeneracy (or number of microstates) $%
W(m)$ of the most probable state $v=m.$ Then 
\begin{equation}
Z(\beta )=e^{-\beta U(m)+\ln W(m)}=W(m)e^{-\beta U(m)}=P(m)  \label{finger}
\end{equation}
is equal to the Gibbs probability of the most probable state $m.$ From (\ref
{thumb}) and (\ref{finger}), we deduce that maximal (minimal) values $F(m)$
of the total free energy give the dominant contribution to $Z(\beta )$ for
negative (resp. positive) temperature.

\smallskip 

\textbf{Remark 0: }\textit{From the proof of the above theorem, it follows
that the most probable state, also the thermodynamically stable state, is
not necessarily the one, call it }$m^{\prime }$\textit{\ that maximizes }$%
W=\exp \left( S/k_B\right) ,$ \textit{but rather that which maximizes the
product }$W\exp (-\beta U).$ \textit{Nonetheless, at temperatures }$T$%
\textit{\ where }$|T|$\textit{\ is small},\textit{\ it is likely that the
state extremizing }$F$\textit{\ is closer to one, call it }$m$\textit{\ that
extremizes }$U$\textit{\ instead of }$m^{\prime }$\textit{\ maximizing }$S;$%
\textit{\ and when }$|T|$\textit{\ is large, it is likely that the state
extremizing }$F$\textit{\ is closer to }$m^{\prime }$\textit{\ that
maximizes }$S$ \textit{than that extremizing }$U.$ \textit{We}\textbf{\ }%
\textit{note the significant point that for systems that support negative
temperatures such as barotropic flows in particular, the specific heat is
nonetheless positive even when the temperature is negative.\medskip
}

A precise correspondence exists between the statistical equilibrium
properties of barotropic flows on one hand and its dynamical properties on
the other. This correspondence is encapsulated in the Minimum Enstrophy or
its equivalent, the extremal kinetic energy principle, which form a pair of
dual variational principles for the steady-states of the barotropic flows
(cf. Leith \cite{Leith85}, Young \cite{Young}, Prieto and Schubert \cite{PS}
and Lim \cite{physfluids05}). The maximal kinetic energy steady-state of the
barotropic flows is closely related to the most probable flow state at hot
enough negative temperatures in the corresponding statistical mechanics
model. Likewise, the minimal kinetic energy steady-state of the barotropic
flows is related to the most probable flow state at low positive
temperatures. At temperatures sufficiently close to $0,$ this correspondence
follows directly from the form of the Gibbs free energy $f=U-TS\simeq U,$ no
matter which of the two general categories below, the barotropic flows
thermal system happens to be in.

In general, the Gibbs canonical ensemble consists of the standard form for
the probability measure 
\[
P_G(w)=\frac 1{Z[\beta ,\mu ]}\exp \left( -\beta H[w]-\mu \Gamma [w]\right) 
\]
where $H[w]$ is the energy and $\beta $ an inverse temperature while $\Gamma
[w]$ represents a significant conserved quantity and $\mu $ a chemical
potential or Lagrange Multiplier conjugate to $\Gamma [w].$ The partition
function or configurational integral $Z[\beta ,\mu ]$ provides the
normalization required to make $P_G[w]$ a probability measure. It is
customary to include only the key conserved quantities as canonical
constraints $\Gamma [w].$ Indeed, in the case of quasi-2D turbulence, there
are an infinite number of conserved quantities, namely the moments of the
relative vorticity or higher order enstrophies, and it is unnecessary to
include all of them as canonical constraints in the Gibbs probability $P_G.$

\section{Spin-Lattice Approximation}

Given the well known fact that Gibbs' canonical ensemble and the
corresponding partition function for the spherical model - to be discussed
in detail below - are closely related to path-integrals and therefore
extremely complex mathematical objects, a rational approximation procedure
based on finite dimensional spin-lattice models or something similar, will
have to be devised to simulate their critical phenomenology on the computer
as well as to solve them by analytical means. Such a rational approximation
scheme must satisfy two basic requirements when the size or order of the
approximation is taken to infinity: (A) the resulting family of finite
dimensional models converge to the correct energy functional and constraints
of the problem and (B) the thermodynamic limit - in this case, the
nonextensive continuum limit - of this family of approximate models exists.
(A) is shown to be true for the family of spin-lattice models given next.
(B) turns out to be true if exact solutions to the spherical models - to be
obtained by the Kac-Berlin method of steepest descent in \cite{Limsphere06}
- yield valid free energy expressions in terms of the associated saddle
points in the nonextensive continuum limit. For the purpose of this paper,
the assumption of the validity of (B) is subsumed under the earlier
assumption that the mean field is asymptotically exact in this class of
problems for the coupled barotropic flows on a rotating sphere.

Using a uniform mesh $M$ of $N$ points $\{x_1,....,x_N\}$ on $S^2$ and the
Voronoi cells based on this mesh - see Lim and Nebus \cite{limnebus} - we
approximate the relative vorticity by 
\[
w(x)=\sum_{j=1}^Ns_jH_j(x) 
\]
where 
\[
s_j=w(x_j)\in (-\infty ,\infty ) 
\]
and $H_j$ is the indicator function on the Voronoi cell $D_j$ centered at $%
x_j,$ that is, 
\begin{eqnarray*}
H_j(x) &=&1\text{ for }x\in D_j, \\
H_j(x) &=&0\text{ for }x\notin D_j.
\end{eqnarray*}

The real valued spins $s_j$ should henceforth be viewed as coarse-grained or
block averaged vorticity - resulting from a single step renormalization
procedure outlined in \cite{Raj} - that is ideally suited to the mean field
approach in this paper. Another point of view describing the real valued
spin states $\{s_j\}$ as macrostates can be found in the book \cite
{limnebusbook}.

Recall that the Voronoi cell $D_j$ is defined to be all the points in $S^2$
that are nearer to $x_j$ than to any other points in the mesh $M.$ This
confers the essential property that the Voronoi cells is a disjoint cover
for $S^2,$ that is,

\begin{eqnarray*}
D_j\cap D_k &=&\emptyset , \\
\bigcup_{j=1}^N &&D_j=S^2.
\end{eqnarray*}
Uniformity of the mesh $M$ confers the other essential property that the
areas $A_j$ of the cells are equal, that is,

\[
A_j=|D_j|=\frac{4\pi }N. 
\]

The rest frame kinetic energy given in terms of the parameters of a frame
that is rotating at the fixed angular velocity of the solid sphere has been
shown to be 
\begin{eqnarray*}
H &=&-\frac 12\int_{S^2}q\psi dx \\
&=&-\frac 12\int_{S^2}dx\left( w+2\Omega \cos \theta \right) G(w)
\end{eqnarray*}
where the fundamental solution of the Laplace-Beltrami operator on $S^2$ is 
\[
\psi (x)=G(w)=\int_{S^2}dx\text{ }\ln |1-x\cdot x^{\prime }|\text{ }%
w(x^{\prime }). 
\]
Under the above approximation for the relative vorticity $w,$ the truncated
or lattice approximate energy 
\begin{eqnarray}
H_N &=&-\frac 12\int_{S^2}dx\left( \sum_{j=1}^Ns_jH_j(x)+2\Omega \cos \theta
\right) G\left( \sum_{j=1}^Ns_jH_j(x)\right)  \nonumber \\
&=&-\frac 12\sum_{j=1}^N\sum_{k=1}^N\left[ \int_{S^2}dx\text{ }H_j(x)G\left(
H_k\right) \right] s_js_k-\Omega \sum_{j=1}^N\left[ \int_{S^2}dx\text{ }\cos
\theta G\left( H_j\right) \right] \text{ }s_j  \label{hn1} \\
&\rightarrow &H\text{ as }N\rightarrow \infty  \nonumber
\end{eqnarray}
under the calculus rules of (Lebesque) integration. With the interactions 
\[
J_{jk}=\int_{S^2}dx\text{ }H_j(x)G\left( H_k\right) 
\]
and the external fields 
\[
F_j=\Omega \int_{S^2}dx\text{ }\cos \theta G\left( H_j\right) , 
\]
the truncated energy takes the standard form of a spin lattice model
Hamiltonian, 
\begin{equation}
H_N=-\frac 12\sum_{j=1}^N\sum_{k=1}^NJ_{jk}s_js_k-\sum_{j=1}^NF_j\text{ }s_j.
\label{hn2}
\end{equation}
The interactions $J_{jk}$ are logarithmic and thus, long range. The external
fields $F_j$ are non-uniform and linear in the spin $\Omega \geq 0,$ and
represent the coupling between the local relative vorticity or spin $s_j$
and the planetary vorticity field $p\cos \theta .$ They are turned off for
the special case of single layer inviscid vortex dynamics on a non-rotating
sphere. The presence of this inhomogeneous term when $\Omega >0,$ is the
source of the much richer mathematical and physical properties of the
coupled barotropic flows on a rotating sphere.

Evaluating the integral and using the essential properties of $H_j$ we
obtain 
\begin{eqnarray}
J_{jk} &=&\int_{S^2}dx\text{ }H_j(x)\int_{S^2}dx^{\prime }\ln |1-x\cdot
x^{\prime }|H_k(x^{\prime })  \nonumber \\
&\rightarrow &\frac{16\pi ^2}{N^2}\ln |1-x_j\cdot x_k|\text{ as }%
N\rightarrow \infty .  \label{jk}
\end{eqnarray}
For the integral in the external fields $F_j$ we obtain 
\begin{eqnarray*}
F_j &=&\Omega \int_{S^2}dx\text{ }\cos \theta \int_{S^2}dx^{\prime }\ln
|1-x\cdot x^{\prime }|H_j(x^{\prime }) \\
&=&\Omega \int_{S^2}dx^{\prime }H_j(x^{\prime })\int_{S^2}dx\text{ }\cos
\theta \ln |1-x\cdot x^{\prime }|
\end{eqnarray*}
after using the symmetry of the inverse $G$ to the Laplace-Beltrami operator
on $S^2.$ Then 
\begin{eqnarray}
F_j &=&\Omega ||\cos \theta ||_2\int_{S^2}dx^{\prime }H_j(x^{\prime
})\int_{S^2}dx\text{ }\psi _{10}(x)\ln |1-x\cdot x^{\prime }|  \nonumber \\
&=&-\frac 12\Omega ||\cos \theta ||_2\int_{S^2}dx^{\prime }H_j(x^{\prime
})\psi _{10}(x^{\prime })  \nonumber \\
&\rightarrow &-\frac{2\pi }N\Omega ||\cos \theta ||_2\psi _{10}(x_j)\text{
as }N\rightarrow \infty ,  \label{fj}
\end{eqnarray}
where $||\cos \theta ||_2$ is the $L_2$ norm of the function $\cos \theta ,$
that is, 
\[
||\cos \theta ||_2=\sqrt{\int_{S^2}dx\text{ }\cos ^2\theta }, 
\]
and the spherical harmonic $\psi _{10}$ which represents the relative
vorticity of solid-body rotation, satisfies the eigenvalue relation, 
\begin{eqnarray}
\psi _{10}(x) &=&\frac{\cos \theta }{||\cos \theta ||_2}=-2G(\psi _{10})
\label{dur} \\
\lambda _{lm} &=&-l(l+1)=-2\text{ for }l=1,m=0.  \nonumber
\end{eqnarray}
The truncated relative enstrophy is given by

\begin{eqnarray}
\Gamma _N &=&\int_{S^2}dx\text{ }w^2=\int_{S^2}dx\left(
\sum_{j=1}^Ns_jH_j(x)\right) ^2  \nonumber \\
&=&\frac{4\pi }N\sum_{j=1}^Ns_j^2\rightarrow \Gamma _r\text{ as }%
N\rightarrow \infty ,  \label{relens}
\end{eqnarray}
after using the two essential properties of the indicator functions $H_j.$

Lastly, the truncated total circulation is given by 
\begin{eqnarray*}
TC_N &=&\int_{S^2}dx\text{ }w=\int_{S^2}dx\text{ }\sum_{j=1}^Ns_jH_j(x) \\
&=&\text{ }\frac{4\pi }N\sum_{j=1}^Ns_j\rightarrow TC\text{ as }N\rightarrow
\infty .
\end{eqnarray*}

\section{Classical and Recent Energy-Enstrophy Theories}

To put the mean field theory here in perspective, we briefly review the two
main theories - both of which are based on Kac's inventions circa 1952 -
used to investigate the statistical relationships between energy and
enstrophy in macroscopic flows. The first is the Gaussian model which forms
the basis of all the classical energy-enstrophy theories. The second is the
spherical model for which the canonical enstrophy constraint is replaced by
the microcanonical form. Versions of the spherical models for new energy -
enstrophy -circulation models of macroscopic flows were first introduced by
the author beginning in early 2000 - see the new Springer-Verlag book by Lim
and Nebus which will be publsihed in October 2006 \cite{limnebusbook}.

\subsection{Gaussian model}

Almost all the early papers on variations of this model used a spectral
formulation in terms of a truncated set of orthnormal eigenfunctions for the
flow domain. Many authors presented results on the applications of this
theory to a large range of topics in geophysical flows including two-layer
flows over nontrivial bottom topography, and quasi-gestrophic f-plane and
beta plane flows \cite{Salmon}. We will use a spatial lattice formulation
instead for reasons already mentioned in the introduction. It is easy to
show that this classical energy-enstropy theory is identical to the
well-known Gaussian Model introduced by Kac \cite{Kac} which is
exactly-solvable.

The classical energy-enstrophy theory as formulated using a spatial
discretization is now given in terms of the truncated energy $H_N$ (\ref{hn2}%
) and relative enstrophy $\Gamma _N$ (\ref{relens}) by the Gaussian
partition function, 
\[
Z_N=\left( \frac b{2\pi }\right) ^{N/2}\int \left[ \prod ds_j\right] \exp
\left( -b\sum_{j=1}^Ns_j^2\right) \exp \left( -\beta H_N[S;\Omega ]\right) , 
\]
in terms of the spin (vorticity) state 
\begin{eqnarray*}
S &=&\{s_1,....,s_N\}, \\
s_j &\in &(-\infty ,\infty ).
\end{eqnarray*}
In this model, the standard deviation of the above Gaussian is given in
terms of the parameter 
\[
b=\frac{4\pi \mu }N 
\]
where $\mu $ is the chemical potential or Lagrange multiplier associated
with the relative enstrophy constraint. The factor $\left( \frac bN\right)
^{N/2}$ is needed to make $W[S]=\left( \frac b{2\pi }\right) ^{N/2}\exp
\left( -b\sum_{j=1}^Ns_j^2\right) $ a probability distribution.

The exact solution of this Gaussian model where it is well-defined, anchors
many of the previous works on statistical equilibrium in geophysical flows
(cf. Salmon, Holloway and Hendershott \cite{Salmon}, Frederiksen and Sawford 
\cite{FS1} and Carnevale and Frederiksen \cite{Carnevale}, amongst many
others).

It turns out as will be shown next, that the choice of a canonical
constraint on total kinetic energy $H[q]$ and a microcanonical constraint on
the relative enstrophy $\Gamma _r$ gives a spherical sodel formulation which
although difficult to solve analytically, is amenable to numerical
simulations and mean field methods. More important is the fact that the
spherical model is well-defined for all temperatures positive or negative,
while the Gaussian model is not defined for certain temperatures.

\subsection{Spherical model for the coupled barotropic flows}

We need to fix the low temperature defect of the classical energy-enstrophy
theory of the barotropic flows which was mentioned above. We will do this by
replacing the canonical constraint on relative enstrophy by a microcanonical
constraint, which yields a version of Kac's spherical model for the
spin-lattice Hamiltonian $H_N$ (\ref{zeen}). The spherical model formulation
of an equilibrium statistical energy-enstrophy theory for the barotropic
flows on a rotating sphere is based on the spin lattice partition function, 
\begin{equation}
Z_N=\int (\Pi ds_j)\delta (NQ_r-\sum s_j^2)\exp \left[ -\beta H_N\right]
\label{zeen}
\end{equation}
where $H_N,$ $J_{jk}$ and the external fields $F_j$ are given in (\ref{hn1}%
), (\ref{jk}) and (\ref{fj}).

The microcanonical relative enstrophy constraint takes the Laplace integral
form and gives 
\[
Z_N=\frac 1{4\pi i}\int (\Pi ds_j)\int_{a-i\infty }^{a+i\infty }d\eta \exp
\left[ \frac 12\eta NQ_r-\frac 12\left\langle S|K|S\right\rangle \right]
\exp \left[ \beta \sum_{j=1}^NF_j\text{ }s_j\right] 
\]
in terms of the matrix 
\[
K_{jk}=\eta -\beta J_{jk} 
\]
and the spin vector $S=\{s_1,...,s_N\}.$

\section{Mean Field Theory}

The spherical model for the energy-enstrophy theory of the rotating
barotropic flows is difficult to solve but it fixes the above foundational
difficulty of the classical energy-enstrophy theory of the coupled
barotropic flows. We therefore use mean field methods to derive thermal
properties of the spin-lattice Hamiltonians, 
\[
H_N=-\frac 12\sum_{j=1}^N\sum_{k=1}^NJ_{jk}s_js_k-\sum_{j=1}^NF_j\text{ }s_j.
\]
This yields physically significant phase transitions for the new
energy-relative enstrophy theory of the coupled barotropic flows on a
rotating sphere. The main step of this mean field theory is to calculate the
change in free energy per site between two fundamentally different vorticity
states, namely, the \textit{mixed macrostate} where the coarse-grained
vorticity of opposite sign are juxtaposed in such a way that a site has
equal probability of having spin $\pm s_0$ and the \textit{unmixed
macrostate }where the coarse-grained vorticity of opposite sign are
separated into hemispheres on $S^2$ in a way that a site has probability $0$
(resp. probability $1)$ of having spin $s_0$ (resp. $-s_0)$ - depending on
where the given site is located relative to the hemispherical pattern.

Many detailed aspects of the formulation below can be changed without any
loss of accuracy or precision to the stated results. Some of these are the
neighborhood size $z$ which determines the range of the interactions in the
model, and the simplified mean spin values $\pm s_0$. For the
energy-enstrophy spin-lattice models, the actual neighborhood $N(j)$ equals
the set of all other sites in the mesh. However, although this signify a
global interaction in $H_N,$ it is a short range one by the accepted
definition - the integral of interactions over all sites is finite - and
therefore, its mean field properties are not dependent on $z.$ Choosing the
simplified mean spin values to be $\pm s_0$ appears to be a more serious
restriction - changing the real valued spins in the original lattice models $%
H_N$ to binary spins that make the resulting model more of an Ising type.
Again, it is not hard to see - by a single step of spatial renormalization
or block spin averaging - that the binary spin models we use in the ensuing
mean field theory are equivalent to the original.

\subsection{Set-up for coupled barotropic flows}

Since the coarse-grained spin or vorticity $s_0(x)$ satisfies the following
integral relations, 
\begin{eqnarray*}
\Gamma &=&\int_{S^2}dx\text{ }s_0^2(x)=Q \\
\int_{S^2}dx\text{ }s_0(x) &=&0=TC,
\end{eqnarray*}
we assume, for the mean field theory, the existence of a site dependent
volume fraction or probability distribution function $\nu _x(s_0)$ such that 
\begin{equation}
\int_{-M}^Mds_0\nu _x(s_0)=1,  \label{vo}
\end{equation}
where $M$ denotes the limiting values of the coarse-grained spin $s_0(x)$
over $S^2$ and 
\begin{eqnarray}
\int_{S^2}dx\int_{-M}^Mds_0\nu _x(s_0)s_0^2 &=&Q  \label{q} \\
\int_{S^2}dx\int_{-M}^Mds_0\nu _x(s_0)s_0 &=&0.  \label{tc11}
\end{eqnarray}

Next we assume that the mixed state and unmixed mean field state, denoted
respectively by 
\begin{equation}
v=m,n,  \label{vm}
\end{equation}
are characterized by $m$ where the mean spins $s_0(x)$ are independent at
neighboring sites and $n$ where the neighboring spins are correlated, that
is, $s_0(x)=s_0(x^{\prime })\in \{s_0^{\pm }\}$. Equivalently, $m$ is
characterized by 
\begin{equation}
\nu _x^m(\pm s_0)=\frac 12\text{ for all }x\in S^2;  \label{tca}
\end{equation}
and $n$ by 
\begin{eqnarray*}
\nu _{x^{+}}^n(s_0) &=&1\text{ for all }x^{+}\in S^2, \\
\nu _{x^{-}}^n(-s_0) &=&1\text{ for all }x^{-}\in S^2
\end{eqnarray*}
where $x^{\pm }$ are lattice sites in respectively the positive and negative
hemispheres of the unmixed state $n.$

The notation $N(j)$ denotes the neighborhood of site $j,$ that is, all
lattice sites $k$ which are connected to site $j.$ Let $|N(j)|=z$ be the
common size of neighborhoods $N(j)$ or coordination number for the lattice.
By varying $z$ we can model interactions of different range in a variety of
spin-lattice models. Let $\varepsilon $ denote the interaction energy scale
(obtained by averaging $J_{jk}$ over $N(j))$ for $H_N.$ For the purpose of
modelling the energy-enstrophy theories, $\varepsilon <0$ which represents
an anti-ferromagnetic interaction.

In addition to the spin-lattice Hamiltonians $H_N,$ we will need a crude
lattice approximation for the per site mixing entropy. Following the same
approach for $J_{jk}$ and $F_j$ in (\ref{jk}) and (\ref{fj}) respectively,
we derive the lattice approximation for the total mixing entropy in the
ensuing steps, 
\begin{eqnarray*}
S[\nu ] &=&-k_B\int_{S^2}dx\text{ }\int_{-M}^Mds\nu _x(s)\ln \nu _x(s) \\
&\simeq &-k_B\sum_{j=1}^N\int_{S^2}dx\text{ }H_j(x)\int_{-M}^Mds\nu _x(s)\ln
\nu _x(s) \\
&=&-\frac{4\pi k_B}N\sum_{j=1}^N\int_{-M}^Mds\nu _{x_j}(s)\ln \nu _{x_j}(s).
\end{eqnarray*}

\subsection{Proofs and results when the solid sphere does not rotate}

First we compute the entropy per site for the mixed state $v=m,$ 
\begin{equation}
S_m=-\frac{4\pi k_B}N\int_{-M}^Mds\nu _x\ln \nu _x=\frac{4\pi k_B}N\ln 2.
\label{5}
\end{equation}
Similarly, the per site entropy in the unmixed state $v=n$ is given by 
\begin{equation}
S_n=0,  \nonumber
\end{equation}
because by the defining property of the unmixed state $v=n,$ neighboring
mean values $s_0(x)$ are correlated perfectly.

Next we compute the mixed state internal energy per site, 
\begin{eqnarray}
u_m &=&-\frac \varepsilon 2\int_{-M}^Mds\nu _x(s)\int_{-M}^Mds^{^{\prime
}}\nu _{x^{\prime }}(s^{^{\prime }})\text{ }s\text{ }s^{\prime }\text{ }z 
\nonumber \\
&=&-\frac{\varepsilon z}2\left[ \int_{-M}^Mds\nu _x(s)\text{ }s\right] ^2 
\nonumber \\
&=&0  \label{five}
\end{eqnarray}
by virtue of property (\ref{tca}). We compute the unmixed state per site
internal energy by using the property that in the unmixed state, neighboring
mean spins $s(x_k)$ satisfy 
\begin{eqnarray*}
s(x_k) &=&s(x_j)=\pm s_0\text{ for all }k\in N(j) \\
\text{with }|N(j)| &=&z\text{ for all }j=1,...,N.
\end{eqnarray*}
Then in this mean field theory, 
\begin{eqnarray}
u_n &=&-\frac{\varepsilon z}{2N}\int_{S^2}dx\int_{-M}^Mds_0\nu _x\text{ }%
s_0^2=-\frac{\varepsilon z}2\int_{-M}^Mds_0\nu _x\text{ }s_0^2  \nonumber \\
&=&-\frac{\varepsilon z}{8\pi }Q  \label{3}
\end{eqnarray}
after using (\ref{q}).

Given the enstrophy $Q>0$ and temperature $T$, and treating the mean spin
distribution $\nu _x$ and state $v=m,n$ as free parameters, the isothermal
free energy difference per site between the mixed and unmixed states is
given by 
\begin{eqnarray}
\Delta f &=&f_m-f_n  \nonumber \\
&=&(u_m-u_n)-T(S_m-S_n)  \nonumber \\
&=&\frac{\varepsilon z}{8\pi }Q-\frac{4\pi k_BT}N\ln 2.  \label{free}
\end{eqnarray}

Equivalently, 
\begin{eqnarray}
f_m &=&u_m-TS_m  \nonumber \\
&=&-\frac{4\pi k_BT}N\ln 2;  \label{freem}
\end{eqnarray}
and 
\begin{eqnarray}
f_n &=&u_n-TS_n  \nonumber \\
&=&-\frac{\varepsilon z}{8\pi }Q.  \label{freen}
\end{eqnarray}

We have nearly finished proving the following mean field result for the
non-rotating barotropic flows using the extension of Planck's theorem that
for negative temperatures $T<0,$ thermodynamically stable statistical
equilibria corresponds to maximizers of the free energy:\smallskip\ 

\textbf{Theorem 5: }\textit{If }$\varepsilon <0$\textit{, then (i) for all }$%
T>0,$\textit{\ the mixed state }$v=m$\textit{\ is preferred. For }$%
\varepsilon <0$\textit{\ and }$T<0,$\textit{\ there is a finite }$N$\textit{%
\ critical temperature } 
\begin{equation}
T_c(N)=\frac{\varepsilon zN}{8\pi S_{ms}}Q<0  \label{tcplu}
\end{equation}
\textit{such that if (ii) } 
\[
0>T>T_c(N)
\]
\textit{then }$v=n$\textit{\ is preferred}$,$\textit{and (iii) if } 
\[
T<T_c(N)<0,
\]
\textit{then }$v=m$\textit{\ is preferred with maximum mean spin entropy }
\begin{equation}
S_{ms}=4\pi k_B\ln 2.  \label{maxen}
\end{equation}

\textit{(iv) In the nonextensive continuum limit - assumed to exist- as }$N$%
\textit{\ }$\rightarrow \infty ,T_c(N)$\textit{\ tends to a finite negative
critical temperature } 
\begin{eqnarray*}
T_c(Q) &=&\frac{\frac Q2\int_{S^2}dx\psi _{10}(x)\int_{S^2}dx^{\prime }\text{
}\psi _{10}(x^{\prime })\ln |1-x\cdot x^{\prime }|}{S_{ms}} \\
&=&\frac{\frac Q2\int_{S^2}dx\psi _{10}(x)G\left( \psi _{10}\right) \text{ }%
(x)}{S_{ms}} \\
&=&\frac{-\frac Q4\int_{S^2}dx\psi _{10}^2(x)}{S_{ms}}=-\frac Q{4S_{ms}}<0
\end{eqnarray*}
\textit{such that if } 
\[
0>T>T_c 
\]
\textit{then }$v=n$\textit{\ is preferred and if } 
\[
T<T_c<0, 
\]
\textit{then }$v=m$\textit{\ is preferred.}

\smallskip\ 

Proof: The proof of (i) follows from the above calculations. To prove cases
(ii) and (iii), we compare the per site free energy 
\begin{equation}
f_m(\max )=-\frac TNS_{ms}>0  \label{min1}
\end{equation}
with the same in the unmixed state $v=n,$ 
\begin{equation}
f_n(\max )=-\frac{\varepsilon z}{8\pi }Q>0.  \label{min2}
\end{equation}
When

\[
-\frac TNS_{ms}<-\frac{\varepsilon z}{8\pi }Q\text{ } 
\]
or equivalently $0>T>T_c(N),$ (ii) $v=n$ is preferred$,$ and vice-versa for
(iii). Since $\varepsilon <0$, $Q>0$ and by definition, $S_{ms}>0,$ the
critical temperature $T_c(N)<0.$

The proof of (iv) follows from consideration of the nonextensive continuum
limit under which the free energy 
\begin{eqnarray*}
\frac{\varepsilon zN}{8\pi }Q &\rightarrow &\frac Q2\int_{S^2}dx\psi
_{10}(x)\int_{S^2}dx^{\prime }\text{ }\psi _{10}(x^{\prime })\ln |1-x\cdot
x^{\prime }| \\
&=&-\frac Q4<0
\end{eqnarray*}
as the number $N$ of lattice sites tend to $\infty ,$ together with the fact
that the denominator in $T_c(N)$ - total entropy $S_{ms}$ in (\ref{maxen}) -
does not depend on $N$.

\smallskip\ 

\textbf{Remark 1}: \textit{When the interaction energy scale }$\varepsilon
<0,$\textit{\ there is a negative temperature transition between the mixed
state at hot }$T<T_c<0$\textit{\ and the unmixed state at very hot }$%
T_c<T<0. $ \smallskip\ 

\textbf{Remark 2}: \textit{The mixed state free energy per site is entirely
entropic. The unmixed state free energy }$f_n$\textit{\ is purely an
internal energy term which is linear in the enstrophy }$Q.$\smallskip 

\subsection{Mean field theory when the solid sphere rotates}

Let us denote the two parts of $H_N$ by 
\begin{eqnarray*}
H_N &=&H_N^{(1)}+H_N^{(2)} \\
H_N^{(1)} &=&-\frac 12\sum_{j=1}^N\sum_{k=1}^NJ_{jk}s_js_k \\
H_N^{(2)} &=&\frac{2\pi \Omega }N\sum_{j=1}^N\text{ }s_j\cos \theta _j.
\end{eqnarray*}
Then we write the corresponding two parts of the internal energies, 
\begin{eqnarray*}
u_m &=&u_m^{(1)}+u_m^{(2)} \\
u_n &=&u_n^{(1)}+u_n^{(2)}
\end{eqnarray*}
with 
\begin{eqnarray*}
u_m^{(1)} &=&0 \\
u_n^{(1)} &=&-\frac{\varepsilon z}{8\pi }Q_r
\end{eqnarray*}
by (\ref{five}) and (\ref{3}) respectively.

Next we assume that the mixed state $v=m$ is defined in (\ref{vm}) and the
unmixed mean field states denoted by $v=n,$ $n_u$ $and$ $n_d$ are
characterized by correlated neighboring spins, $s(x)=s(x^{\prime })\in
\{s_0^{\pm }\}.$ Moreover, the unmixed states $v=n,$ $n_u$ and $n_d$ are
both exactly correlated by hemispheres into opposite values $s_0^{\pm }$
such that 
\begin{eqnarray}
s_0^{+}+s_0^{-} &=&0  \label{unun} \\
2\pi \left( (s_0^{+})^2+(s_0^{-})^2\right) &=&Q_r.  \label{dudu}
\end{eqnarray}
Solving for $s_0^{\pm }$ we get 
\begin{equation}
s_0^{+}=-s_0^{-}=\sqrt{\frac{Q_r}{4\pi }}.  \label{spino}
\end{equation}

The unmixed states $v=n_u$ and $n_d$ are further characterized by
hemispherically correlated spins $s_0^{\pm }$ which satisfies the
expressions (\ref{upp}) and (\ref{down}) respectively, that is, aligned and
anti-aligned with the northern and southern hemispheres corresponding to the
planetary spin $\Omega >0.$

The entropy calculations are exactly the same as when planetary spin $\Omega
=0$ because the entropies $S_m$ and $S_n$ depends only on the statistical
distribution $v_x(s_0)$ of the mean field relative vorticity $s_0.$

Next we compute the values of the per site internal energies $u_m^{(2)}$ and 
$u_n^{(2)}$ due to the nonzero planetary spin $\Omega >0$ of $H_N:$ using
definition (\ref{vm}) for the mixed state $v=m,$ we get 
\begin{eqnarray*}
u_m^{(2)} &=&\frac{2\pi \Omega }{N^2}\left\langle \sum_{j=1}^N\text{ }%
s_j\cos \theta _j\right\rangle \\
&=&\frac{2\pi \Omega }{N^2}\sum_{j=1}^N\text{ }\cos \theta _j\left\langle
s_j\right\rangle \\
&=&\frac{2\pi \Omega }{N^2}\left( \int_{-M}^Mds\nu _x(s)\text{ }s\right)
\sum_{j=1}^N\text{ }\cos \theta _j \\
&=&0
\end{eqnarray*}
since (\ref{tca}). Denoting by 
\[
x_j^{\pm },\theta _k^{\pm } 
\]
the lattice sites and co-latitudes that fall respectively into the
hemispheres determined by the correlated means $s_0^{\pm },$ we obtain for
the generic unmixed state $v=n,$ 
\begin{eqnarray*}
u_n^{(2)} &=&\frac{2\pi \Omega }{N^2}\left\langle \sum_{j=1}^N\text{ }%
s_j\cos \theta _j\right\rangle =\frac{2\pi \Omega }{N^2}\sum_{j=1}^N\text{ }%
\left\langle s_j\right\rangle \cos \theta _j \\
&\equiv &\frac{2\pi \Omega }{N^2}\left\{ \left( \sum_{j=1}^{N/2}\text{ }%
\left\langle s(x_j^{+})\right\rangle \cos \theta _j^{+}\right) +\text{ }%
\left( \sum_{k=1}^{N/2}\text{ }\left\langle s(x_k^{-})\right\rangle \cos
\theta _k^{-}\right) \right\} \\
&=&\frac{2\pi \Omega }{N^2}\left( s_0^{+}\sum_{j=1}^{N/2}\text{ }\cos \theta
_j^{+}+s_0^{-}\sum_{k=1}^{N/2}\text{ }\cos \theta _k^{-}\right)
\end{eqnarray*}
where the hemispheres defined by $s_0^{\pm }$ need not be aligned with the
northern and southern hemispheres corresponding to the spin $\Omega >0.$

Next we compute the per site free energies 
\begin{eqnarray}
f_m &=&u_m-TS_m  \nonumber \\
&=&\frac{4\pi k_BT}N\ln 2  \label{free}
\end{eqnarray}
and 
\begin{eqnarray}
f_n &=&u_n-TS_n  \nonumber \\
&=&-\frac{\varepsilon z}{8\pi }Q_r+\frac{2\pi \Omega }{N^2}\left(
s_0^{+}\sum_{j=1}^{N/2}\text{ }\cos \theta _j^{+}+s_0^{-}\sum_{k=1}^{N/2}%
\text{ }\cos \theta _k^{-}\right) .  \label{freenn}
\end{eqnarray}
Therefore, the per site change in free energy is now given by 
\begin{eqnarray*}
\Delta f &=&f_m-f_n \\
&=&\frac{4\pi k_BT}N\ln 2
\end{eqnarray*}
\[
+\frac{\varepsilon z}{8\pi }Q_r-\frac{2\pi \Omega }{N^2}\left(
s_0^{+}\sum_{j=1}^{N/2}\text{ }\cos \theta _j^{+}+s_0^{-}\sum_{k=1}^{N/2}%
\text{ }\cos \theta _k^{-}\right) . 
\]

\smallskip\ 

\textbf{Remark 3: }\textit{Application of this formulation to the
non-rotating barotropic flows on the sphere at both positive and negative
temperatures requires consideration of the nonextensive continuum limit
under which } 
\begin{eqnarray*}
\varepsilon z &=&\varepsilon (N)z(N)\rightarrow O(N^{-1})<0, \\
\frac{2\pi }{N^2}\sum_{j=1}^{N/2} &&\text{ }s_0^{+}\cos \theta _j^{+}+\frac{%
2\Omega }{N^2}\sum_{k=1}^{N/2}\text{ }s_0^{-}\cos \theta _k^{-}\rightarrow
O(N^{-1}).
\end{eqnarray*}

\smallskip\ 

\subsubsection{Positive temperature}

For $T>0,$ we will need to compare the minimum free energy per site in the
mixed state $v=m,$ 
\begin{equation}
f_m(\min )=-\frac TNS_{ms}=\frac{4\pi k_BT}N\ln 2  \label{min2}
\end{equation}
with the same in the unmixed state $v=n,$ 
\begin{equation}
f_n(\min )=-\frac{\varepsilon z}{8\pi }Q_r+\frac{2\pi \Omega }{N^2}\left(
s_0^{+}\sum_{j=1}^{N/2}\text{ }\cos \theta _j^{+}+s_0^{-}\sum_{k=1}^{N/2}%
\text{ }s_0^{-}\cos \theta _k^{-}\right) .  \label{min22}
\end{equation}
The extreme value of $f_n(\min )$ is obtained at the most negative value of 
\[
\frac{2\pi \Omega }{N^2}\left( s_0^{+}\sum_{j=1}^{N/2}\text{ }\cos \theta
_j^{+}+s_0^{-}\sum_{k=1}^{N/2}\text{ }\cos \theta _k^{-}\right) 
\]
which occurs when the hemispheres associated with the correlated means $%
s_0^{\pm }$ are anti-correlated with those corresponding to the planetary
spin $\Omega >0.$

For any $T>0,$ if 
\begin{equation}
\frac{2\pi \Omega }{N^2}\left( s_0^{+}\sum_{j=1}^{N/2}\text{ }\cos \theta
_j^{+}+s_0^{-}\sum_{k=1}^{N/2}\text{ }\cos \theta _k^{-}\right) >\frac{%
\varepsilon z}{8\pi }Q_r-\frac TNS_{ms}  \label{rhs}
\end{equation}
then $v=m$ is preferred; if the inequality is reversed then an unmixed state
is preferred.

Thus, if $\varepsilon <0,$ then for all $T>0$ and $Q_r>0,$ the RHS of (\ref
{rhs}) is negative$.$ This implies that for all $T>0$, $Q_r>0,$ and $\Omega
>0,$ the mixed state $v=m$ is preferred over any $v=n_u$ state where the
hemispheres associated with the means $s_0^{\pm }$ are correlated with those
corresponding to spin $\Omega >0,$ i.e., 
\begin{equation}
\frac 1N\left( s_0^{+}\sum_{j=1}^{N/2}\text{ }\cos \theta
_j^{+}+s_0^{-}\sum_{k=1}^{N/2}\text{ }\cos \theta _k^{-}\right) >0.
\label{upp}
\end{equation}

Next, using (\ref{rhs}), we compare the mixed state $v=m$ with the unmixed
states $v=n_d$ where the hemispheres associated with the means $s_0^{\pm }$
are anti-correlated with those corresponding to spin $\Omega >0,i.e.,$%
\begin{equation}
\frac 1N\left( \text{ }s_0^{+}\sum_{j=1}^{N/2}\cos \theta
_j^{+}+s_0^{-}\sum_{k=1}^{N/2}\text{ }\cos \theta _k^{-}\right) <0.
\label{down}
\end{equation}
For $\varepsilon <0$ and any fixed relative enstrophy $Q_r>0,$ there is a
positive finite size $N$ critical temperature depending on $\Omega $ and $%
Q_r,$%
\begin{equation}
T_c(\Omega ,Q_r;N)=\frac{2\pi \Omega I_{-}-\frac{\varepsilon z}{8\pi }NQ_r}{%
-S_{ms}(Q_r)}>0  \label{tcc}
\end{equation}
where 
\begin{equation}
-\infty <I_{-}\equiv \min_{s_0^{\pm }}\frac 1N\left( s_0^{+}\sum_{j=1}^{N/2}%
\text{ }\cos \theta _j^{+}+s_0^{-}\sum_{k=1}^{N/2}\text{ }\cos \theta
_k^{-}\right) <0,  \label{iminus}
\end{equation}
provided the spin $\Omega $ is large enough relative to $Q_r,$ that is, 
\begin{equation}
\Omega >\Omega _{+}(Q_r)=\frac{\varepsilon z}{16\pi ^2I_{-}}NQ_r
\label{spplus}
\end{equation}
which converges to a finite limit as $N\rightarrow \infty .$ The minimum in $%
I_{-}$ (\ref{iminus}) is taken over all possible orientations of the
hemispheres associated with $s_0^{\pm }$, and by (\ref{spino}), 
\[
I_{-}=O(\sqrt{Q_r}). 
\]

In summary, (a) for $T>T_c(N)>0,$ the mixed state $v=m$ is preferred over
any unmixed state $v=n_u$, $n_d$ and (b) for $0<T<T_c(N),$ the unmixed state 
$v=n_d$ satisfying (\ref{down}) is preferred over the mixed state $v=m;$ (b)
in the nonextensive continuum limit as $N\rightarrow \infty $, the positive
finite size $N$ critical temperature 
\[
T_c(\Omega ,Q_r;N)\rightarrow T_c(\Omega ,Q_r)<\infty 
\]
and moreover, if 
\[
\Omega >\Omega _{+}(Q_r)=\frac{\varepsilon z}{16\pi ^2I_{-}}NQ_r 
\]
holds, then 
\[
T_c(\Omega ,Q_r)>0. 
\]

\smallskip\ 

\textbf{Remark 4: }\textit{Thus, there is a positive temperature phase
transition if the planetary spin is large enough.}\smallskip\ 

\subsubsection{Negative temperature}

An extension of Planck's theorem to negative temperature states that for $%
T<0,$ the state with maximum free energy is preferred. Thus, we compare the
per site free energies of the mixed $v=m$ and unmixed states $v=n_u$ and $%
n_d:$%
\begin{eqnarray*}
f_m(\max ) &=&-\frac TNS_{ms}, \\
f_n(\max ) &=&-\frac{\varepsilon z}{8\pi }Q_r+\frac{2\pi \Omega }{N^2}\left(
s_0^{+}\sum_{j=1}^{N/2}\text{ }\cos \theta _j^{+}+s_0^{-}\sum_{k=1}^{N/2}%
\text{ }\cos \theta _k^{-}\right) .
\end{eqnarray*}

For $T<0,$ 
\begin{equation}
-\frac{\varepsilon z}{8\pi }Q_r+\frac{2\pi \Omega }{N^2}\left(
s_0^{+}\sum_{j=1}^{N/2}\text{ }\cos \theta _j^{+}+s_0^{-}\sum_{k=1}^{N/2}%
\text{ }\cos \theta _k^{-}\right) >-\frac TNS_{ms}>0  \label{compmin}
\end{equation}
implies that the unmixed states are preferred. Solving this inequality for
the finite size $N$ critical temperature we get 
\begin{equation}
T_c^{-}(\Omega ,Q_r;N)=\frac{2\pi \Omega I_{+}-\frac{\varepsilon z}{8\pi }%
NQ_r}{-S_{ms}}<0  \label{tcm}
\end{equation}
where 
\begin{equation}
\infty >I_{+}\equiv \max_{s_0^{\pm }}\frac 1N\left( s_0^{+}\sum_{j=1}^{N/2}%
\text{ }\cos \theta _j^{+}+s_0^{-}\sum_{k=1}^{N/2}\text{ }\cos \theta
_k^{-}\right) =-I_{-}>0  \label{iplus}
\end{equation}
and 
\[
I_{+}=O(\sqrt{Q_r}). 
\]
In this case when (c) $T<T_c^{-}(N)<0,$ the mixed state $v=m$ is preferred
over any aligned unmixed state $v=n_u$ and thus, over any anti-aligned
unmixed state $v=n_d$ as well since the left hand side of (\ref{compmin})
evaluated at $v=n_d$ and $v=n_u$ satisfies 
\[
\left[ -\frac{\varepsilon z}{8\pi }Q_r+\frac{2\pi \Omega }{N^2}\left(
s_0^{+}\sum_{j=1}^{N/2}\text{ }\cos \theta _j^{+}+s_0^{-}\sum_{k=1}^{N/2}%
\text{ }\cos \theta _k^{-}\right) \right] (n_d)< 
\]
\[
\left[ -\frac{\varepsilon z}{8\pi }Q_r+\frac{2\pi \Omega }{N^2}\left(
s_0^{+}\sum_{j=1}^{N/2}\text{ }\cos \theta _j^{+}+s_0^{-}\sum_{k=1}^{N/2}%
\text{ }\cos \theta _k^{-}\right) \right] (n_u)<-\frac TNS_{ms}. 
\]
When (d) $T_c^{-}(N)<T<0,$ the aligned unmixed state $v=n_u$ is preferred
over $v=m.$ It is important to note that a negative finite size $N$ critical
temperature $T_c^{-}(\Omega ,Q_r;N)<0$ exists for any spin $\Omega >0,$
unlike the positive finite size critical temperature $T_c(N)$ which exists
only for spins satisfying (\ref{spplus}).\smallskip\ 

\textbf{Remark 5: }\textit{Under the nonextensive continuum limit }$%
N\rightarrow \infty ,$\textit{\ } 
\begin{eqnarray*}
T_c^{-}(\Omega ,Q_r;N) &\rightarrow &T_c^{-}(\Omega ,Q_r)<\infty \\
T_c^{-}(\Omega ,Q_r) &<&0,
\end{eqnarray*}
\textit{because in (\ref{tcm}), the denominator is negative by definition
and does not depend on }$N,$\textit{\ and the numerator tends to a finite
positive limit by virtue of Remark 3. This implies that there is a negative
temperature phase transition for all values of the planetary spin.\smallskip%
\ }

Since the numerator in (\ref{tcm}) contains the positive term $-\frac{%
\varepsilon z}{8\pi }NQ_r$ (for $\varepsilon <0),$ we can define a second
negative critical temperature 
\begin{equation}
T_c^{--}(\Omega ,Q_r;N)=\frac{2\pi \Omega I_{-}-\frac{\varepsilon z}{8\pi }%
NQ_r}{(-S_{ms})}<0  \label{tcmm}
\end{equation}
provided the spin $\Omega >0$ is small enough, that is, 
\begin{equation}
\Omega <\Omega _{+}(Q_r)=\frac{\varepsilon z}{16\pi ^2I_{-}}NQ_r.
\label{spmin}
\end{equation}
We note that $T_c^{--}(\Omega ,Q_r;N)$ $<0$ and $T_c(N)>0$ in (\ref{tcc})
are the same expression corresponding to the two sides of the equality in (%
\ref{spmin}). In this case when (e) $T<T_c^{--}(N)<0,$ the mixed state $v=m$
is preferred over the anti-aligned unmixed state $v=n_d$ and when (f) $%
T_c^{--}(N)<T<0,$ the anti-aligned unmixed state $v=n_d$ is preferred over $%
v=m.$

Comparing (\ref{tcm}) to (\ref{tcmm}), we deduce that when the spin $\Omega $
satisfies (\ref{spmin}), 
\[
T_c^{-}(N)<T_c^{--}(N)<0. 
\]
By comparing the per site free energies of unmixed states $v=n_u$ and $n_d$
when $T<0$ (more specifically $T_c^{-}<T<T_c^{--}<0$ and $%
T_c^{-}<T_c^{--}<T<0)$, 
\begin{eqnarray*}
f_n^u(\max ) &=&-\frac{\varepsilon z}{8\pi }Q_r+\frac{2\pi \Omega }NI_{+} \\
f_n^d(\max ) &=&-\frac{\varepsilon z}{8\pi }Q_r+\frac{2\pi \Omega }NI_{-},
\end{eqnarray*}
we deduce at once that the aligned unmixed state $v=n_u$ is always preferred
over the anti-aligned unmixed state $v=n_d$ for $T<0.$

\smallskip\ 

\textbf{Remark 6: }\textit{Under the nonextensive continuum limit }$%
N\rightarrow \infty ,$\textit{\ } 
\[
T_c^{--}(N)\rightarrow T_c^{--}<\infty 
\]
\textit{and } 
\[
T_c^{-}<T_c^{--}<0 
\]
\textit{provided the planetary spin is small enough, that is, } 
\[
\Omega <\Omega _{+}(Q_r)=\frac{\varepsilon z}{16\pi ^2I_{-}}NQ_r. 
\]

\smallskip\ 

Collecting together the above computations, we summarize the phase
transitions in this simple mean field theory for the coupled barotropic
flows - rotating sphere system:\smallskip\ 

\textbf{Theorem 6: }\textit{(A) for large enough spins }$\Omega >0,$\textit{%
\ the anti-aligned unmixed state }$v=n_d$\textit{\ changes into the mixed
state }$v=m$\textit{\ at }$T_c(\Omega ,Q_r)>0,$\textit{\ (B) this mixed
state }$v=m$\textit{\ continues to be preferred for all positive }$T>T_c$%
\textit{\ and negative }$T<T_c^{-},$\textit{\ changing to the aligned
unmixed state }$v=n_u$\textit{\ at }$T_c^{-}(\Omega ,Q_r)<0$\textit{, and
(C) (i) for large enough spins }$\Omega ,$\textit{\ the unmixed state }$%
v=n_u $\textit{\ persists for all }$T$\textit{\ such that }$T_c^{-}<T<0,$%
\textit{\ and (ii) for small enough spins }$\Omega <\Omega _{+}(Q_r),$%
\textit{\ the state }$v=n_u$\textit{\ persists as the preferred state for
all }$T$\textit{\ such that }$T_c^{-}<T<0,$\textit{\ but for }$T$\textit{\
such that }$T_c^{--}<T<0,$\textit{\ the state }$v=n_d$\textit{\ is preferred
over }$v=m$\textit{\ but not }$v=n_u.$

\textit{In the nonextensive continuum limit, the mean field critical
temperatures of the energy-relative enstrophy theory for this problem}%
\textbf{: } 
\begin{eqnarray*}
\infty &>&T_c(\Omega ,Q_r)=\lim_{N\rightarrow \infty }\frac{2\pi \Omega
I_{-}-\frac{\varepsilon z}{8\pi }NQ_r}{(-S_{ms})} \\
&=&\frac{\min_{w,Q_r}\left( \int_{S^2}dx\text{ }\Omega C\psi
_{10}G[w(x^{\prime })]\right) \text{ }+\frac{Q_r}4}{(-S_{ms})} \\
&=&\frac{\frac 12\Omega C\sqrt{Q_r}\text{ }-\frac{Q_r}4}{S_{ms}}>0
\end{eqnarray*}
\textit{\ precisely} \textit{when} 
\begin{eqnarray*}
\Omega &>&\Omega _{+}^\infty (Q_r)=\frac{\text{ }\sqrt{Q_r}}{\text{ }2C} \\
&>&0;
\end{eqnarray*}
\textit{and for all planetary spins }$\Omega >0,$%
\begin{eqnarray*}
-\infty &<&T_c^{-}(\Omega ,Q_r)=\lim_{N\rightarrow \infty }\frac{2\pi \Omega
I_{+}-\frac{\varepsilon z}{8\pi }NQ_r}{(-S_{ms})} \\
&=&\frac{\max_{w,Q_r}\left( \int_{S^2}dx\text{ }\Omega \cos \theta
(x)G[w(x^{\prime })]\right) \text{ }+\frac{Q_r}4}{(-S_{ms})} \\
&=&\frac{-\frac 12\Omega C\sqrt{Q_r}\text{ }-\frac{Q_r}4}{S_{ms}}<0,
\end{eqnarray*}
\textit{where }$I^{\pm \text{ }}$ are given by (\ref{iplus}) and (\ref
{iminus}), $C=\sqrt{\int_{S^2}dx\text{ }\cos ^2\theta },$ the $\min_{w,Q_r}$
(resp. $\max_{w,Q_r})$ is taken over all relative vorticity $w(x)$ with
fixed relative enstrophy $Q_r$ and the total entropy is \textit{\ } 
\[
S_{ms}\equiv -4\pi k_B\ln 2>0. 
\]
\smallskip\ 

Proof: Using the definitions of the interaction energy scale $\varepsilon
(N) $ and the size $z(N)$ of the interaction neighborhood, we deduce that
the term $-\frac{\varepsilon z}{8\pi }NQ_r$ in the numerator of $T_c(N)$,
being the finite dimensional representation of the per site spin-spin
interaction internal energy in the unmixed states $v=n_u$ or $n_d$ (as
opposed to the spin-$\Omega $ interaction term$),$tends to 
\begin{equation}
\text{ }-\frac{Q_r}2\int_{S^2}dx\psi _{10}(x)\int_{S^2}dx^{\prime }\psi
_{10}(x^{\prime })\ln |1-x\cdot x^{\prime }|=\frac{Q_r}4>0.  \label{pooh}
\end{equation}

Thus, from the definition of the minimum $I_{-}$, and the derivation of the
spin-lattice Hamiltonians $H_N$ from the rest frame pseudo kinetic energy $H$
of the coupled barotropic flows, the finite $N$ critical temperature 
\[
T_c(\Omega ,Q_r;N)=\frac{2\pi \Omega \min_{s_0^{\pm }}\frac 1N\left(
s_0^{+}\sum_{j=1}^{N/2}\text{ }\cos \theta _j^{+}+s_0^{-}\sum_{k=1}^{N/2}%
\text{ }\cos \theta _k^{-}\right) -\frac{\varepsilon z}{8\pi }NQ_r}{(-S_{ms})%
} 
\]
\[
\rightarrow T_c(\Omega ,Q_r)=\frac{\min_{w,Q_r}\left( \int_{S^2}dx\text{ }%
\Omega \cos \theta (x)G[w(x^{\prime })]\right) \text{ }-\frac{Q_r}4}{%
(-S_{ms})}. 
\]
where the denominator does not depend on $N.$ The minimum in the numerator
is taken over all relative vorticity $w(x)$ with fixed relative enstrophy $%
Q_r,$ and is a well-defined finite negative quantity that is proportional to 
$\Omega $ and to $\sqrt{Q_r}$ - 
\begin{eqnarray*}
\min_{w,Q_r}\left( \int_{S^2}dx\text{ }\Omega C\psi _{10}G[w(x^{\prime
})]\right) &=&\int_{S^2}dx\text{ }\Omega C\psi _{10}G[\sqrt{Q_r}\psi _{10}]
\\
&=&-\frac 12\Omega C\sqrt{Q_r}\int_{S^2}dx\text{ }\psi _{10}^2 \\
&=&-\frac 12\Omega C\sqrt{Q_r}
\end{eqnarray*}
Since 
\begin{eqnarray*}
\Omega _{+}(Q_r) &=&\frac{\varepsilon z}{16\pi ^2I_{-}}NQ_r \\
&\rightarrow &\Omega _{+}^\infty (Q_r)=\frac{-Q_r}{\min_{w,Q_r}\left(
\int_{S^2}dx\text{ }4C\psi _{10}G[w(x)]\right) }>0,
\end{eqnarray*}
when $\Omega >\Omega _{+}^\infty (Q_r),$ the mean field critical temperature 
\[
T_c(\Omega ,Q_r)>0. 
\]

A similar argument based on 
\begin{eqnarray*}
\max_{w,Q_r}\left( \int_{S^2}dx\text{ }\Omega C\psi _{10}G[w(x^{\prime
})]\right) &=&\int_{S^2}dx\text{ }\Omega C\psi _{10}G[-\sqrt{Q_r}\psi _{10}]
\\
&=&\frac 12\Omega C\sqrt{Q_r}
\end{eqnarray*}
proves the existence of the continuum limit of $T_c^{-}$ to conclude the
proof.

\smallskip\ 

It is useful to restate and separate the above phase transitions into two
categories depending on planetary spin.

\smallskip\ 

\textbf{Theorem 7: }\textit{When the planetary spin }$\Omega <\Omega
_{+}(Q_r),$\textit{\ (i) there is no positive temperature phase transition
and the mixed state }$v=m$\textit{\ is preferred for all }$T>0,$\textit{\
(ii) there is a negative temperature phase transition at }$T_c^{-}<0$\textit{%
\ (which exists irrespective of the value of planetary spin), where the
preferred mixed state }$v=m$\textit{\ for all }$T<T_c^{-}<0,$\textit{\
changes into the aligned state }$v=n_u$\textit{\ which is preferred over
both }$v=m$\textit{\ and }$v=n_d$\textit{\ for all }$T_c^{-}<T<0,$\textit{\
and (iii) there is a secondary transition at the hotter temperature }$%
T_c^{--}<0$\textit{\ (that is, }$T_c^{-}<T_c^{--}),$\textit{\ where the
intermediate state }$v=n_d$\textit{\ changes place with }$v=m$\textit{\ in
order of thermal preference. Letting }$\prec $\textit{\ denote `has smaller
free energy than', we summarize the state preference for the case }$\Omega
<\Omega _{+}(Q_r):$\textit{\ } 
\begin{eqnarray*}
n_d &\prec &n_u\prec m\text{ for }T<T_c^{-}<0 \\
n_d &\prec &m\prec n_u\text{ for }T_c^{-}<T<T_c^{--} \\
m &\prec &n_d\prec n_u\text{ for }T_c^{--}<T<0.
\end{eqnarray*}

\textit{When the planetary spin }$\Omega >\Omega _{+}(Q_r),$\textit{\ (iv)
there is a positive critical temperature }$T_c(\Omega ,Q_r)$\textit{\ given
by (\ref{tcc}) at which the preferred state changes from }$v=n_d$\textit{\
for }$0<T<T_c$\textit{\ to }$v=m$\textit{\ for all }$T>T_c$\textit{\ and all 
}$\func{negative}$\textit{\ }$T<T_c^{-}<0,$\textit{\ and (v) there is a
negative critical temperature }$T_c^{-}<0$\textit{\ that exists irrespective
of the value of planetary spin, at which the preferred state changes from }$%
v=m$\textit{\ to }$v=n_u$\textit{\ for all negative }$T>T_c^{-}.$\textit{\
We summarize the state preference for the case }$\Omega >\Omega _{+}(Q_r):$%
\textit{\ } 
\begin{eqnarray*}
n_d &\prec &n_u\prec m\text{ for }T<T_c^{-}<0 \\
n_d &\prec &m\prec n_u\text{ for }T_c^{-}<T<0.
\end{eqnarray*}

\section{Conclusion}

A related approach used to derive nonlinear stability properties of
steady-states in the coupled barotropic flow - rotating sphere system is a
variational formulation that is based on extremizing the rest frame kinetic
energy of flow for fixed relative enstrophy \cite{Limsiap}, \cite{Lim05a}, 
\cite{physfluids05}. This approach is the dual of the Minimum Enstrophy
method where enstrophy is extremized under fixed kinetic energy without
fixing angular momentum. The Minimum Enstrophy Method is in turn related to
the dynamical asymptotic principle of Selective Decay for damped 2D
turbulence which states that the quotient of enstrophy to energy tends to a
minimum in time. Amongst many others, the Minimum Enstrophy methods with
fixed angular momentum were first used by Leith \cite{Leith85}, Young \cite
{Young} and Prieto and Schubert \cite{PS} in geophysical flows. The main
result in \cite{physfluids05} gives precise necessary and sufficient
conditions for the existence and nonlinear stability of prograde and
retrograde solid-body flows in the coupled barotropic fluid - rotating solid
sphere system in terms of the planetary spin, relative enstrophy and rest
frame kinetic energy. These conditions have been useful also as starting
points in Monte-Carlo simulations of the spherical model for barotropic
flows in \cite{Ding} and will be compared to the results of the mean field
theory here as well as the results in \cite{Raj}.

Results from all three methods are closely correlated with recent rigorous
results on the nonlinear stability of steady-states of barotropic flows \cite
{Lim05a}, \cite{Limsiap}, \cite{physfluids05} and results from direct
numerical simulations of a pde closely related to the coupled barotropic
fluid - rotating solid sphere system \cite{Cho}. The relationship between
nonlinear stability and statisical equilibrium properties of geophysical
flows is however, not new. Carnevale and Frederiksen discussed this
relationship in their paper \cite{Carnevale}, and Shepherd et al \cite
{shepherd90}, \cite{shepherdmu1}, \cite{shepherdwho} are good references on
applications of the energy-Casimir method in geophysical flows. In
particular, Shepherd et al \cite{shepherd}, \cite{Wshepherd} have found
convincing arguments for zonal anisotropy in rapidly rotating barotropic
flows (and the lack of such in slowly or non-rotating flows), in the sense
that, the Arnold stability theorems used to prove the nonlinear stability of
zonal basic flows, fail when the planetary spin is small.

In conclusion and comparison with the extensive Monte-Carlo simulation
results on the same problems (cf. Ding and Lim \cite{Ding}), the numerical
asymptotic results for the dissipative coupled barotropic fluid - rotating
solid sphere system in Cho and Polvani \cite{Cho} and the variational
results in Lim \cite{physfluids05}, we observe that \textit{(1) the mixed
state }$v=m$\textit{\ is preferred for all positive temperature except when
the planetary spin is large compared to relative enstrophy, whence the
counter-rotating unmixed state }$v=n_d$\textit{\ arises at low enough
positive temperatures, (2) the mixed state }$v=m$\textit{\ is preferred for
negative temperatures that are more negative than }$T_c^{-}<0$\textit{\
(when the system has intermediate kinetic energies) while the pro-rotating
state }$v=n_u$\textit{\ is preferred for negative temperatures hotter or
less negative than }$T_c^{-}$\textit{\ (when the system has extremely high
kinetic energies), (3a) the positive critical temperature} $T_c$ \textit{%
increases with planetary spin }$\Omega $ \textit{and (3b) the negative
critical temperature }$T_c^{-}$\textit{\ decreases linearly in the planetary
spin }$\Omega .$\textit{\ Its dependence on relative enstrophy is more
complex, being a sum of two terms, one of which is linear in }$Q_r$\textit{\
and other is proportional to }$\sqrt{Q_r}$\textit{; an increase in }$Q_r$%
\textit{\ results in a decrease in }$T_c^{-}$\textit{, that is, }$T_c^{-}$%
\textit{\ becomes more negative and less hot.}

Given the accepted temperature convention that the more negative $T<0$ is
less hot (or less energetic), and negative temperatures are hotter (or more
energetic) than positive ones, we note that the Monte-Carlo results of Ding
and Lim \cite{Ding}, the dynamic simulation results of Cho and Polvani \cite
{Cho} and the variational steady state results of Lim \cite{physfluids05}
are all consistent with the mean field results in this paper. Moreover, they
complement each other.

In particular, when $T$ is negative and hotter than $T_c^{-}(\Omega ,Q_r)$
(or have relatively high kinetic energy), the prograde solid-body state $%
v=n_u$ is preferred over both the mixed state $v=m$ and the retrograde
solid-body state $v=n_d,$ according to result (2) in our mean field theory;
according to simulations \cite{Ding} of the statistical equilibrium at very
hot negative temperatures, the prograde solid-body flow is the most probable
vortex state; according to the numerical results on the dissipative
barotropic flows in Cho (cf. fig. 8 in \cite{Cho}), there is a robust
relaxation to the prograde solid-body state when the planetary spin $\Omega $
is not too large for a given range of flow kinetic energy, which is
consistent with result (3b) that $T_c^{-}$\textit{\ decreases linearly in
the planetary spin }$\Omega ;$ and finally according to the steady-state /
nonlinear stability results in Lim \cite{physfluids05}, the prograde
solid-body state is allowed only when the planetary spin is small compared
to the kinetic energy, and in this case, it is nonlinearly stable (and hence
observable).

According to result (1) in our mean field theory, the mixed vortex state $%
v=m $ is preferred for all positive temperatures, unless planetary spin is
large compared to relative enstrophy, in which case, the retrograde
solid-body state $v=n_d$ is preferred over both the mixed state and the
prograde solid-body state $v=n_u$ when positive $T$ is lower than $%
T_c(\Omega ,Q_r).$ According to the MC-simulator results in Ding and Lim 
\cite{Ding}, the statistical equilibria of the barotropic flows relaxes to
the retrograde solid-body state for relatively low positive temperatures and
large planetary spins; this is consistent with our result (3a) that $T_c$%
\textit{\ increases with the planetary spin }$\Omega .$ According to the
numerical results on the dissipative coupled barotropic fluid - rotating
solid sphere system in Cho (cf. fig. 9 in \cite{Cho}), the asymptotic state
reached when planetary spin is relatively large, is dominated by large
anti-cyclonic polar vortices; allowing for the fact that the large polar
vortex can be expressed in the form of a superposition of mainly the
retrograde solid-body state and some zonally symmetric spherical harmonics
of low order, their numerical result is partially consistent with our
results (1) and (3a). Finally, according to the steady-state / nonlinear
stability results in Lim \cite{physfluids05}, the retrograde solid-body
state arises for all values of the planetary spin and relative enstrophy,
but it is nonlinearly stable (and hence observable) only when the planetary
spin is large compared to relative enstrophy; this is consistent with our
mean field results.

\section{Appendix: Non-divergent vs divergent flows}

The non-divergent case of the coupled barotropic fluid - rotating solid
sphere system in this paper is much simpler to treat than the divergent case
which is of course more realistic. This more realistic model is essentially
based on coupling the shallow-water equations (coupledSWE) model to a
rotating solid sphere through a complex torque mechanism that allows the
divergent fluid in this case to exchange energy as well as angular momentum
with the sphere. We will summarize aspects of the coupledSWE in order to
state explicitly the approximations in the coupled barotropic fluid -
rotating solid sphere system. For this purpose let us denote by $U,$ $L$ and 
$H,$ the velocity, length and depth scales respectively. Then two important
dimensionless numbers are the Rossby and Froude numbers respectively, 
\[
R=\frac U{2\Omega L},\text{ }F=\frac U{\sqrt{gH}}, 
\]
where $g$ is the gravitational constant. Within the coupled SWE model, the
relative importance of convective / local flow to rotational effects is
measured by the Rossby number $R.$ The Froude number $F$ measures the
relative importance within the coupled SWE model of the convective / local
flow effects to gravity-depth effects. In a definite sense, a small Rossby
number $R\ll 1$ signals the importance of rotational effects: $\Omega $ has
to be relatively large or the scale $L$ of the flow has to be relatively
large in order for rotation of the planet to be important. On the other
hand, a large Froude number $F$ $\gg 1$ implies the importance of gravity
effects, since in this case, the gravity waves are relatively slow, and
cannot be time-averaged out of the problem.

The Rossby Radius of deformation, 
\[
L_R=\frac{\sqrt{gH}}{2\Omega }=\frac{RL}F, 
\]
measures the relative importance of gravity-depth effects to rotational
effects. When it is of $O(1),$ both gravity and rotational effects are
relevant to the problem, and only when $L_R\gg 1$ that rotational and
convective or inertial effects dominate. It is convenient to label the
square of $L_R/L$ by the Burgers number 
\[
B=\frac{R^2}{F^2}=\frac{L_R^2}{L^2}. 
\]
Small values of $B$ signals the importance of gravity-depth effects over
rotational effects; it inludes the case when the Rossby number $R$ itself is
relatively small, that is, when rotational effects dominate convective or
inertial effects, as well as the case when $R$ is $O(1)$, that is, when
rotational effects are comparable to convective or inertial effects.

The coupled barotropic fluid - rotating solid sphere system can be
characterized as the limit of the coupled SWE when $L_R$ tends to $\infty $
or equivalently when the depth scale $H$ tends to $\infty $ with $\Omega $
and $g$ fixed. The flow $(u,v)$ in the coupled barotropic fluid - rotating
solid sphere system model is strictly 2D, that is, $\omega $ is a scalar
field and the top and bottom boundary conditions are idealized away by
taking, in effect, the depth scale $H$ to $\infty $. Thus, in a definite
sense, the coupled barotropic fluid - rotating solid sphere system models a
rotating fluid of infinite depth. This fact partly accounts for its
tractability relative to the more complex coupled SWE model where boundary
conditions at the top and bottom of the fluid are retained. In this definite
sense, the coupled barotropic fluid - rotating solid sphere system model is
non-divergent, i.e., $div$ $(u,v)=0,$ while the coupled SWE is a divergent
model.

\smallskip\ \textsc{Acknowledgement}

This work is supported by ARO grant W911NF-05-1-0001 and DOE grant
DE-FG02-04ER25616; the author would like to acknowledge the scientific
support of Drs. Chris Arney, Anil Deane, Gary Johnson and Robert Launer.

\end{document}